\documentclass{article}
\title{Derivations for Locating Photon Emission Points Using Compton Imaging in GRETINA}
\author{Dr. Robert Crabbs \\ University of California, Berkeley \\
	\and 
	Dr. I-Yang Lee \\ Lawrence-Berkeley National Laboratory \\
	\and 
	Dr. Kai Vetter \\	Lawrence-Berkeley National Laboratory \\
	}
\date{\today}

\usepackage{amsmath} 
\usepackage{bm} 
\usepackage{graphicx} 
\usepackage{pdfpages} 
\usepackage{textcomp} 
\usepackage{color} 
\usepackage{multicol,caption} 
\usepackage{geometry} 
\usepackage{hyperref} 

\hypersetup{
  colorlinks   = true, 
  urlcolor     = blue, 
  linkcolor    = black, 
  citecolor   = red 
}
\newcommand{\norm}[1]{\lVert#1\rVert}

\newenvironment{multicolfigure}
  {\par\medskip\noindent\minipage{\linewidth}}
  {\endminipage\par\medskip}

\begin{document}

\maketitle

\begin{abstract}
GRETA, the \textbf{G}amma-\textbf{R}ay \textbf{E}nergy \textbf{T}racking \textbf{A}rray, is an array of highly-segmented HPGe detectors designed to track $ \gamma $-rays emitted in beam-physics experiments. Its high detection efficiency and state-of-the-art position resolution make it well-suited for imaging applications. In this paper, we derive the expressions for locating a photon emission point using Compton imaging. We also include expressions for corresponding uncertainty calculations.
\end{abstract}



\section{Introduction}
\label{sec:introduction}

In a typical in-beam experiment, a projectile nucleus is accelerated to high energy and directed into a beam target. Nuclear reactions in the target produce recoil nuclei heading downstream. These recoil nuclei are often excited, and can emit one or more characteristic $ \gamma $-ray photons somewhere downstream of the target. By locating the emission points of these photons, we can determine the lifetime of the excited recoil nuclei.
\par
Compton imaging is one method by which we can locate photon emissions. \cite{compton_imaging_in_gretina} A $ \gamma $-ray typically interacts several times in an HPGe detector before being fully-absorbed, resulting in a sequence of hits in the detector denoted $ \bm{X_1} $ to $ \bm{X_N} $. (Figure \ref{fig:compton_imaging_geometry}) However, we cannot directly measure the interaction sequence because the detector electronics are not fast enough to resolve the differences in timing. Instead, we use Compton sequencing to deduce the sequence. \cite{compton_sequencing_in_gretina}
\par

\begin{multicols}{2}
Once we have the interaction sequence, we can define a ``Compton cone'' from detected energy depositions and the locations of the first 2 interactions. Each cone shows the possible directions from which a photon came as it entered the detector, and is uniquely defined by its vertex $ \bm{X_C} $, central axis $ \bm{\hat{V}_C} $, and cosine of opening angle $ \mu_C = \cos{\theta_C} $.
\par
The intersections of these Compton cones with the recoil beam are the possible emission points of the photons. Note that a cone will intersect a beam at (up to) 2 points. In practice, our cones will intersect the beam at 2 distinct locations, or else not at all. Real-world detector position \& energy resolution can cause errors in the locations, energy depositions, and sequence of interactions; such errors can significantly skew the resulting Compton cones. The following sections describe the math behind finding an emission point, $ \bm{X_0} $.

\begin{multicolfigure}
\centering
\includegraphics[width=0.95\textwidth]{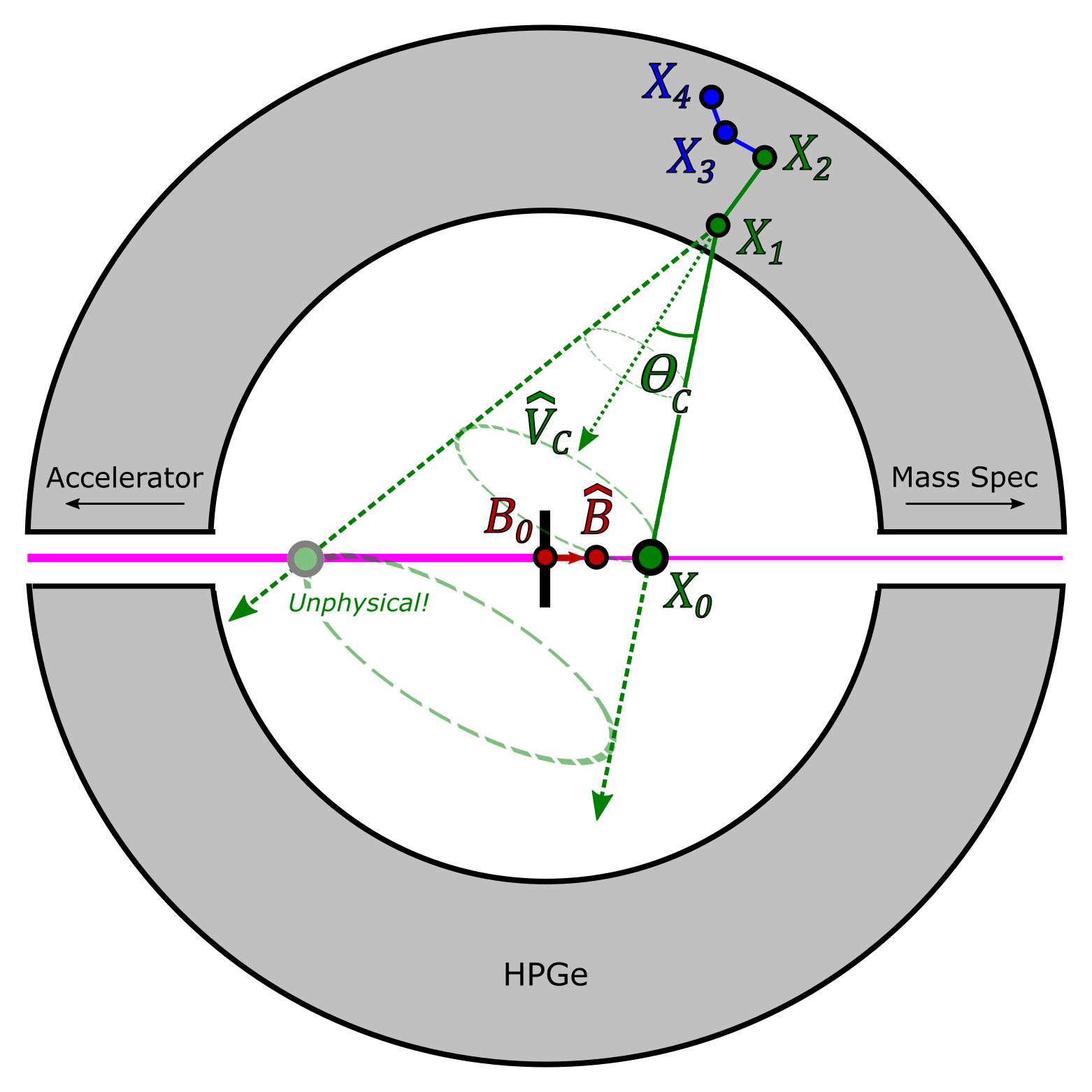}
\captionof{figure}{Geometry Used in Compton Imaging}
\label{fig:compton_imaging_geometry}
\end{multicolfigure}

\end{multicols}


\section{Cone-Beam Intersections}
\label{sec:cone_beam_intersections}

Reference \cite{intersection_line_cone} provides a general geometric expression for the intersections between cones \& beams, which we can adapt here for our specific problem. We seek an emission point $ \bm{X_0} $ somewhere on the beam. Let's start by parametrizing our recoil beam as $ \bm{B}(t) = \bm{B_0} + \bm{\hat{B}} t $, with beam axis $ \bm{\hat{B}} $ and beam anchor point $ \bm{B_0} $. In GRETINA or GRETA, the beamline lies along the $ \bm{\hat{z}} $-axis and passes through the origin of the lab frame, so we can simply set $ \bm{B_0} = (0,0,0) $ and $ \bm{\hat{B}} = (0,0,1) $. This simplifies our expression for the emission point to $ \bm{X_0} = \bm{B}(t) = \bm{\hat{B}} t $. 
\par
As noted in Section \ref{sec:introduction}, a Compton cone is defined by its vertex $ \bm{X_C} $, central axis $ \bm{\hat{V}_C} $, and cosine of opening angle $ \mu_C = \cos{\theta_C} $. Our emission point $ \bm{X_0} $ must also lie somewhere on this cone, which is true if:
\begin{align}
\label{eq:point_on_a_cone1}
\bm{\hat{V}_C} \cdot \frac{(\bm{X_0} - \bm{X_C})}{\norm{\bm{X_0} - \bm{X_C}}} = \mu_C
\end{align}
Re-arranging terms, we can rewrite this equation as:
\begin{align}
\label{eq:point_on_a_cone2}
\bm{\hat{V}_C} \cdot (\bm{X_0} - \bm{X_C}) = \mu_C \norm{\bm{X_0} - \bm{X_C}}
\end{align}
Squaring both sides, we get:
\begin{align}
\label{eq:point_on_a_cone3}
(\bm{\hat{V}_C} \cdot (\bm{X_0} - \bm{X_C}))^2 = \mu_C^2 \norm{\bm{X_0} - \bm{X_C}}^2
\end{align}
Using the vector identities $ (\bm{a} \cdot (\bm{b} - \bm{c}))^2 = (\bm{b} - \bm{c})^T \bm{a} \bm{a}^T (\bm{b} - \bm{c}) $ and $ (\bm{b} - \bm{c})^T (\bm{b} - \bm{c}) = \norm{\bm{b} - \bm{c}}^2 $, the above becomes:
\begin{align}
\label{eq:point_on_a_cone4}
 (\bm{X_0} - \bm{X_C})^T \bm{\hat{V}_C} \bm{\hat{V}_C}^T (\bm{X_0} - \bm{X_C}) = \mu_C^2 (\bm{X_0} - \bm{X_C})^T (\bm{X_0} - \bm{X_C})
\end{align}
\begin{align}
\label{eq:point_on_a_cone5}
 (\bm{X_0} - \bm{X_C})^T M (\bm{X_0} - \bm{X_C}) = 0
\end{align}
where $ M = \bm{\hat{V}_C} \bm{\hat{V}_C}^T - \mu_C^2 I_3 $ is a 3 $ \times $ 3 symmetric matrix. Substituting $ \bm{X_0}(t) = \bm{\hat{B}} t $ in Equation \ref{eq:point_on_a_cone5}, we get:
\begin{align}
\label{eq:point_on_a_cone6}
 (\bm{\hat{B}} t - \bm{X_C})^T M (\bm{\hat{B}} t - \bm{X_C}) = 0
\end{align}
\begin{align}
\label{eq:point_on_a_cone7}
\bm{\hat{B}}^T M \bm{\hat{B}} t^2 - (\bm{\hat{B}}^T M \bm{X_C} - \bm{X_C}^T M \bm{\hat{B}}) t + \bm{X_C}^T M \bm{X_C} = 0
\end{align}
We can simplify this expression with the matrix identity $ (DEF)^T = F^T E^T D^T $, noting that:
\begin{align}
(\bm{X_C}^T M \bm{\hat{B}})^T = \bm{\hat{B}}^T M^T \bm{X_C} = \bm{\hat{B}}^T M \bm{X_C}
\end{align}
because M is symmetric. Since M is a 3 $ \times $ 3 matrix, we know $ \bm{X_C} $ is a 3 $ \times $ 1 column vector and $ \bm{\hat{B}}^T $ is a 1 $ \times $ 3 row vector. Therefore, $ \bm{\hat{B}}^T M \bm{X_C} $ and $ \bm{X_C}^T M \bm{\hat{B}} $ are both 1 $ \times $ 1 scalars, and transposing them makes no difference. Therefore we can say that: 
\begin{align}
\bm{\hat{B}}^T M \bm{X_C} &= (\bm{X_C}^T M \bm{\hat{B}})^T \\ 
&= \bm{X_C}^T M \bm{\hat{B}}
\end{align}
This fact lets us rewrite Equation \ref{eq:point_on_a_cone7} as a quadratic equation in $ t $:
\begin{align}
\label{eq:point_on_a_cone8}
\bm{\hat{B}}^T M \bm{\hat{B}} ~ t^2 - 2 \bm{\hat{B}}^T M \bm{X_C} ~ t + \bm{X_C}^T M \bm{X_C} = 0
\end{align}
Let $ a = \bm{\hat{B}}^T M \bm{\hat{B}} = \bm{\hat{B}} \cdot (M \bm{\hat{B}}) $, $ b = -2 \bm{\hat{B}}^T M \bm{X_C} = -2 \bm{\hat{B}} \cdot (M \bm{X_C}) $, and $ c = \bm{X_C}^T M \bm{X_C} = \bm{X_C} \cdot (M \bm{X_C}) $, so $ a t^2 + b t + c = 0 $. This can be readily solved for t, and from there the cone-beam intersections are given by $ \bm{X_{0,1}} = t_1 \bm{\hat{B}} $ and $ \bm{X_{0,2}} = t_2 \bm{\hat{B}} $. 
\par
As mentioned before, when using imperfect detectors a Compton cone may not intersect the beamline at all. This is the situation when $ b^2 - 4ac < 0 $.

\section{Error Propagation}
\label{sec:error_propagation_compton_imaging}

We can compute analytical error estimates to judge the reliability of the reconstruction for an individual photon track. The full derivation contains 20 pages of partial derivatives -- those have been omitted and ``left as an exercise to the reader''.
\par
Recall the general error propagation formula for a dependent variable $ z $:
\begin{align}
\label{eq:error_propagation_formula}
\sigma_z^2 = \sum_{j = 1}^{N} \left(\frac{\partial z}{\partial x_j}\right)^2 \sigma_{x_j}^2
\end{align}
where $ (x_1, ..., x_N) $ are the independent variables from which $ z $ is calculated.
\par
In Compton imaging, it might appear that we would only need 3 such variables to define a Compton cone: a vertex, an axis, and an opening angle. However, these 3 variables actually correspond to 8 independent variables. The cone's vertex is the first hit in the photon track: $ \bm{X_C} = \bm{X_1} = (x_1, x_2, x_3) $. With the second photon hit at $ \bm{X_2} = (x_4, x_5, x_6) $, the cone axis is defined:
\begin{align}
\bm{\hat{V}_C} &= (v_1, v_2, v_3) \\
&= \frac{\bm{X_1} - \bm{X_2}}{\norm{\bm{X_1} - \bm{X_2}}} \\
&= \left( \frac{x_1 - x_4}{L}, \frac{x_2 - x_5}{L}, \frac{x_3 - x_6}{L} \right)
\end{align}
where $ L = \sqrt{(x_1 - x_4)^2 + (x_2 - x_5)^2 + (x_3 - x_6)^2} $ is the Compton ``lever arm''. Lastly, the cone angle is determined by:
\begin{align}
\mu_C = 1 - {m_e}c ^2 \left( \frac{1}{E_1} - \frac{1}{E_0} \right)
\end{align}
where $ E_0 $ is the photon's lab-frame emission energy and $ E_1 $ is the energy after the initial scatter in the detector. In all, we therefore need to measure 8 independent quantities to obtain a Compton reconstruction: $ (x_1, x_2, x_3, x_4, x_5, x_6, E_0, E_1) $. 
\par
Note that in a real-world experiment, $ \bm{\hat{B}}$ and $ \bm{B_0} $ are not constant. Finite beam spot sizes and straggling can cause minor variations in the energies and directions of recoil nuclei. To account for such variations in the recoil beam, then, we would also have to add the photon \textit{parent}'s trajectory to the list of independent variables. The velocity component is 4 variables: $ \beta \bm{\hat{B}} = (\beta b_{1}, \beta b_{2}, \beta b_{3}) $, where $ \bm{\hat{B}} $ is the heading for a particular recoil nucleus and $ \beta c $ is its speed. We would also need a beam setpoint, $ \bm{B_{0}} = (b_{0,1}, b_{0,2}, b_{0,3}) $ that the recoil nucleus passes through. In all, this would add another 7 independent variables to the analysis: $ (b_1, b_2, b_3, b_{0,1}, b_{0,2}, b_{0,3}, \beta) $. For simplicity, though, and because the variations in the recoil beam are typically not large, we have chosen to ignore these variables here. 
\par
In Section \ref{sec:cone_beam_intersections} we went over the math behind finding emission points with Compton imaging. The goal was to calculate the $ t $'s in $ \bm{X_0} = \bm{B_0} + t\bm{\hat{B}} $. In the end, we arrived at a quadratic equation:
\begin{align}
\label{quadratic_equation_for_t}
t &= \frac{-b \pm \sqrt{b^2 - 4 a c}}{2a} \\
a &= \bm{\hat{B}} \cdot (M \bm{\hat{B}}) \\
b &= -2 \bm{\hat{B}} \cdot (M \bm{X_C}) \\
c &= \bm{X_C} \cdot (M \bm{X_C})
\end{align}
Again, $ M = \bm{\hat{V}_C} \bm{\hat{V}_C}^T - \mu_C^2 I_3 $ is a 3 $ \times $ 3 symmetric matrix:
\begin{align}
M = \left( \begin{array}{ccc}
v_1^2 - \mu_C^2 & v_1 v_2 & v_1 v_3 \\
v_1 v_2 & v_2^2 - \mu_C^2 & v_2 v_3 \\
v_1 v_3 & v_2 v_3 & v_3^2 - \mu_C^2
\end{array} \right)
\end{align}
To get scalar expressions for error analysis, we evaluate $ a $, $ b $, and $ c $ with known quantities for the beam: 
\begin{align}
\bm{B_0} = (0,0,0) & & \bm{\hat{B}} = (0,0,1)
\end{align}
The matrix algebra is straightforward, yielding the following:
\begin{align}
a &= v_3^2 - \mu_C^2 \\
b &= -2 [ v_1 v_3 x_1 + v_2 v_3 x_2 + (v_3^2 - \mu_C^2) x_3 ] \\
\begin{split}
c {}&= x_1^2 (v_1^2 - \mu_C^2) + x_2^2 (v_2^2 - \mu_C^2) + x_3^2 (v_3^2 - \mu_C^2) \\
& \quad + 2 ( v_1 v_2 x_1 x_2 + v_1 v_3 x_1 x_3 + v_2 v_3 x_2 x_3 )
\end{split}
\end{align}
Furthermore, note that $ \bm{X_0} = \bm{B_0} + t\bm{\hat{B}} = (0,0,t) $, so we can identify the emission point simply by its z-coordinate, $ z = t $.
\par
Now that we have these expressions, the partial derivatives of z with respect to the independent variables $ (x_1, ..., x_6, E_0, E_1) $ can be calculated using the chain rule:
\begin{equation}
\label{eq:the_chain_rule}
\frac{\partial z(y_1,...,y_N)}{\partial x_j} = \sum_{n = 1}^{N} \left(\frac{\partial z}{\partial y_n}\right) \left(\frac{\partial y_n}{\partial x_j}\right)
\end{equation}
Here, $ z $ is a function of the variables $ a(x_1, ..., x_6, E_0, E_1) $, $ b(x_1, ..., x_6, E_0, E_1) $, and $ c(x_1, ..., x_6, E_0, E_1) $ from our earlier quadratic equation. So, for example:
\begin{equation}
\frac{\partial z}{\partial x_1} = \frac{\partial z}{\partial a} \frac{\partial a}{\partial x_1} + \frac{\partial z}{\partial b} \frac{\partial b}{\partial x_1} + \frac{\partial z}{\partial c} \frac{\partial c}{\partial x_1}
\end{equation}
\par
Calculating the partial derivatives is a lengthy ordeal, and so we'll just provide the results below. Note that the partial derivatives of $ z $ contain a $ \pm $ or $ \mp $ sign, which reflect the root chosen in Equation \ref{quadratic_equation_for_t}. This only works when Compton imaging yields a unique emission point for the photon.
\par
Now that we have partial derivatives for $ z $ with respect to each of the independent variables, we are close to an analytic estimate of the total error. To avoid confusion, we will refer to this estimated imaging resolution as $ \sigma_{img} $ instead of $ \sigma_z $. The next step is to get a measure of uncertainty for each of the independent variables. In practice, GRETINA's position \& energy resolution are both energy-dependent. In addition, position resolution is not spherically-symmetric -- GRETINA's modules give positions more precisely radially than axially. For simplicity, we have assumed that $ (x_1, ..., x_6) $ are all equally-affected by the detector's position resolution, $ \sigma_{xyz} $, and $ E_0 $ and $ E_1 $ are equally-affected by the detector's energy resolution, $ \sigma_E $. Typical values are $ \sigma_{xyz} $ = 3.0 mm and $ \sigma_E $ = 2.0 keV.
\par
Summing everything together with the Chain Rule, we find:
\begin{align}
\begin{split}
\sigma_{img}^2 {}& = \left[ \left( \frac{\partial z}{\partial x_1} \right)^2 + \left( \frac{\partial z}{\partial x_2} \right)^2  + \left( \frac{\partial z}{\partial x_3} \right)^2 + \left( \frac{\partial z}{\partial x_4} \right)^2 + \left( \frac{\partial z}{\partial x_5} \right)^2 + \left( \frac{\partial z}{\partial x_6} \right)^2 \right] \sigma_{xyz}^2 \\
& \quad + \left[ \left( \frac{\partial z}{\partial E_0} \right)^2 + \left( \frac{\partial z}{\partial E_1} \right)^2 \right] \sigma_E^2
\end{split} \\[1.5ex]
& = \sigma_{img,pos}^2 + \sigma_{img,energy}^2
\end{align}
We have separated the resolution estimate into two components -- one depending entirely on position resolution ($ \sigma_{img,pos} $) and one only on energy resolution ($ \sigma_{img,energy} $). For detector resolutions typical of GRETINA, $ \sigma_{img,pos} >> \sigma_{img,energy} $ in Compton imaging.

\newpage
\begin{center}
\textbf{Partial Derivatives of z}
\end{center}
\begin{align}
\frac{\partial z}{\partial a}& = \frac{b}{2 a^2} \pm \frac{c - b^2 / 2 a}{a \sqrt{b^2 - 4 a c}} \\[1.5ex]
\frac{\partial z}{\partial b}& = \frac{-1 \pm b / \sqrt{b^2 - 4 a c}}{2 a} \\[1.5ex]
\frac{\partial z}{\partial c}& = \frac{\mp 1}{\sqrt{b^2 - 4 a c}}
\end{align}
~\\

\begin{center}
\textbf{Partial Derivatives of a}
\end{center}
\begin{align}
\frac{\partial a}{\partial x_1}& = - \frac{2 v_1 v_3^2}{L} &
\frac{\partial a}{\partial x_4}& = - \frac{\partial a}{\partial x_1} \\[1.5ex]
\frac{\partial a}{\partial x_2}& = - \frac{2 v_2 v_3^2}{L} &
\frac{\partial a}{\partial x_5}& = - \frac{\partial a}{\partial x_2} \\[1.5ex]
\frac{\partial a}{\partial x_3}& = \frac{2 v_3 (1 - v_3^2)}{L} &
\frac{\partial a}{\partial x_6}& = - \frac{\partial a}{\partial x_3} \\[1.5ex]
\frac{\partial a}{\partial E_0}& = \frac{2 \mu m_e}{E_0^2} &
\frac{\partial a}{\partial E_0}& = - \frac{2 \mu m_e}{E_1^2}
\end{align}

\begin{center}
\textbf{Partial Derivatives of b}
\end{center}
\begin{align}
\frac{\partial b}{\partial x_1}& = \frac{2 v_3  (2 v_1^2 x_1 - 2 x_1 + x_4) + 4 v_1 v_2 v_3 x_2}{L} - 2 x_3 \frac{\partial a}{\partial x_1} \\[1.5ex]
\frac{\partial b}{\partial x_2}& = \frac{2 v_3  (2 v_2^2 x_2 - 2 x_2 + x_5) + 4 v_1 v_2 v_3 x_1}{L} - 2 x_3 \frac{\partial a}{\partial x_2} \\[1.5ex]
\frac{\partial b}{\partial x_3}& = \frac{2 (2 v_3^2 - 1) (v_1 x_1 + v_2 x_2)}{L} - 2 (a + x_3 \frac{\partial a}{\partial x_1}) \\[1.5ex]
\frac{\partial b}{\partial x_4}& = - \frac{2 v_3 x_1 (2 v_1^2 - 1) - 4 v_1 v_2 v_3 x_2}{L} - 2 x_3 \frac{\partial a}{\partial x_4} \\[1.5ex]
\frac{\partial b}{\partial x_5}& = - \frac{2 v_3 x_2 (2 v_2^2 - 1) - 4 v_1 v_2 v_3 x_1}{L} - 2 x_3 \frac{\partial a}{\partial x_5} \\[1.5ex]
\frac{\partial b}{\partial x_6}& = - \frac{2 (2 v_3^2 - 1) (v_1 x_1 + v_2 x_2)}{L} - 2 x_3 \frac{\partial a}{\partial x_6}
\end{align}
\begin{align}
\frac{\partial b}{\partial E_0} = -2 x_3 \frac{\partial a}{\partial E_0} &&
\frac{\partial b}{\partial E_1} = -2 x_3 \frac{\partial a}{\partial E_1}
\end{align}
\par
\newpage

\begin{center}
\textbf{Partial Derivatives of c}
\end{center}
\begin{align}
\begin{split}
\frac{\partial c}{\partial x_1} = {}& 2 v_1 x_1^2 (1 - v_1^2) / L + 2 v_1 v_2 x_2 (1 - v_2 x_2) / L - 4 v_1 v_2 v_3 x_2 x_3 / L \\
& \quad + 2 x_1 (v_2 x_2 + v_3 x_3)(1 - 2 v_1^2) / L + x_3^2 \frac{\partial a} {\partial x_1} + 2 x_1 (v_1^2 - \mu^2) + 2 v_1 v_3 x_3
\end{split} \\
\begin{split}
\frac{\partial c}{\partial x_2} = {}& 2 v_2 x_2^2 (1 - v_2^2) / L + 2 v_1 v_2 x_1 (1 - v_1 x_1) / L - 4 v_1 v_2 v_3 x_1 x_3 / L \\
& \quad + 2 x_2 (v_1 x_1 + v_3 x_3)(1 - 2 v_2^2) / L + x_3^2 \frac{\partial a} {\partial x_2} + 2 x_2 (v_2^2 - \mu^2) + 2 v_2 v_3 x_3
\end{split} \\
\begin{split}
\frac{\partial c}{\partial x_3} = {}& 2 x_3 (v_1 x_1 + v_2 x_2) (1 - 2 v_3^2) / L - 2 v_3 (v_1^2 x_1^2 + v_2^2 x_2^2) / L - 4 v_1 v_2 v_3 x_1 x_2 / L \\
& \quad + 2 v_3 (v_1 x_1 + v_2 x_2) + x_3^2 \frac{\partial a} {\partial x_3} + 2 x_3 a
\end{split} \\
\begin{split}
\frac{\partial c}{\partial x_4} = {}& -2 v_1 x_1^2 (1 - v_1^2) / L + 2 v_1 v_2^2 x_2^2 / L + 4 v_1 v_2 v_3 x_2 x_3 / L \\
& \quad - 2 x_1 (v_2 x_2 + v_3 x_3)(1 - 2 v_1^2) / L + x_3^2 \frac{\partial a} {\partial x_4}
\end{split} \\
\begin{split}
\frac{\partial c}{\partial x_5} = {}& -2 v_2 x_2^2 (1 - v_2^2) / L + 2 v_2 v_1^2 x_1^2 / L + 4 v_1 v_2 v_3 x_1 x_3 / L \\
& \quad - 2 x_2 (v_1 x_1 + v_3 x_3)(1 - 2 v_2^2) / L + x_3^2 \frac{\partial a} {\partial x_5}
\end{split} \\
\begin{split}
\frac{\partial c}{\partial x_6} = {}& -2 x_3 (v_1 x_1 + v_2 x_2) (1 - 2 v_3^2) / L + 2 v_3 (v_1^2 x_1^2 + v_2^2 x_2^2) / L \\
& \quad + 4 v_1 v_2 v_3 x_1 x_2 / L + x_3^2 \frac{\partial a} {\partial x_6}
\end{split} \\[1.5ex]
\frac{\partial c}{\partial E_0} = {}& \frac{2 \mu m_e}{E_0^2} (x_1^2 + x_2^2) + x_3^2 \frac{\partial a} {\partial E_0} \\[1.5ex]
\frac{\partial c}{\partial E_1} = {}& -\frac{2 \mu m_e}{E_1^2} (x_1^2 + x_2^2) + x_3^2 \frac{\partial a} {\partial E_1}
\end{align}

~\\
\par
As mentioned, the derivation of the above partial derivatives was a lengthy process. We did try Mathematica's procedural derivative-calculator, but the generated formulae  were much less tractable -- sometimes taking up a full page each. Still, we were able to double-check our work numerically. Inputting hypothetical values of $ x_1 $ and the other independent variables produced identical results for both sets of formulae.
\par
We also performed an independent numerical check by tweaking each independent variable for the sample Compton reconstruction in Table \ref{table:compton_error_estimate_check} below. By introducing small changes in $ x_1 $ or another of the variables, one can obtain numerical derivatives to compare with the results of the analytical expressions above. We used 0.01 mm increments for positions and 0.025 keV increments for energies.
\par
\begin{table}[ht]
\centering 
\begin{tabular}{c | c c c c}
\textbf{Hit} & \textbf{x (mm)} & \textbf{y (mm)} & \textbf{z (mm)} & \textbf{$ \Delta E $ (keV)} \\ [0.5ex]
\hline
\#1 & -81.4542 & 172.4690 & -30.0678 & 288.4240 \\
\#2 & -100.3864 & 193.4548 & -49.6538 & 210.0868 \\
\#3 & -101.4564 & 193.9505 & -52.1735 & 474.4475 \\
\#4 & -96.0459 & 197.7831 & -44.1857 & 0.6924 \\
\#5 & -95.9333 & 197.8570 & -43.9780 & 199.5873 \\
\hline
\end{tabular}
\centering
\caption{Sample photon track used in error estimates}
Hits: 5, Track Energy: 1173.238keV
\end{table}

\begin{table}[ht]
\centering 
\begin{tabular}{c c c || c c c}
\textbf{Partial} & \textbf{Analytic} & \textbf{Numeric} & \textbf{Partial} & \textbf{Analytic} & \textbf{Numeric} \\
\textbf{Derivative} & \textbf{Estimate} & \textbf{Estimate} & \textbf{Derivative} & \textbf{Estimate} & \textbf{Estimate} \\ [0.5ex]
\hline
$ \partial z/\partial a $ & -0.08191 & -- & $ \partial \theta/\partial E_0 $ & 0.000723 & 0.000720 \\
$ \partial z/\partial b $ & 0.02344 & -- & $ \partial \theta/\partial E_1 $ & -0.001271 & -0.001266 \\
$ \partial z/\partial c $ & -0.00671 & -- & -- & -- & -- \\ \hline
$ \partial a/\partial x_1 $ & -0.01039 & -0.01040 & $ \partial b/\partial x_1 $ & -4.0290 & -4.0284 \\
$ \partial a/\partial x_2 $ & 0.01152 & 0.01150 & $ \partial b/\partial x_2 $ & 1.7436 & 1.7467 \\
$ \partial a/\partial x_3 $ & 0.02238 & 0.02237 & $ \partial b/\partial x_3 $ & 5.2351 & 5.2424 \\
$ \partial a/\partial x_4 $ & 0.01039 & 0.01040 & $ \partial b/\partial x_4 $ & 3.4018 & 3.4033 \\
$ \partial a/\partial x_5 $ & -0.01152 & -0.01153 & $ \partial b/\partial x_5 $ & -1.0484 & -1.0458 \\
$ \partial a/\partial x_6 $ & -0.02244 & -0.02237 & $ \partial b/\partial x_6 $ & -4.4115 & -4.4043 \\
$ \partial a/\partial E_0 $ & 0.000637 & 0.000640 & $ \partial b/\partial E_0 $ & 0.0383 & 0.0383 \\
$ \partial a/\partial E_1 $ & -0.001120 & -0.001120 & $ \partial b/\partial E_1 $ & -0.0674 & -0.0673 \\ \hline
$ \partial c/\partial x_1 $ & -167.46 & -166.95 & $ \partial z/\partial x_1 $ & 1.0295 & 1.0253 \\
$ \partial c/\partial x_2 $ & -734.68 & -733.81 & $ \partial z/\partial x_2 $ & 4.9671 & 4.9613 \\
$ \partial c/\partial x_3 $ & -780.21 & -780.22 & $ \partial z/\partial x_3 $ & 5.3535 & 5.3571 \\
$ \partial c/\partial x_4 $ & 103.26 & 103.73 & $ \partial z/\partial x_4 $ & -0.6136 & -0.6170 \\
$ \partial c/\partial x_5 $ & 684.85 & 685.43 & $ \partial z/\partial x_5 $ & -4.6166 & -4.6217 \\
$ \partial c/\partial x_6 $ & 633.98 & 633.96 & $ \partial z/\partial x_6 $ & -4.3535 & -4.3510 \\
$ \partial c/\partial E_0 $ & 23.752 & 23.746 & $ \partial z/\partial E_0 $ & -0.1585 & -0.1584 \\
$ \partial c/\partial E_1 $ & -41.761 & -41.755 & $ \partial z/\partial E_1 $ & 0.2786 & 0.2785 \\
\hline
\end{tabular}
\caption{Numerical check of Compton imaging error estimates}
\label{table:compton_error_estimate_check}
~\\
\end{table}

\begin{table}[ht]
\centering 
\begin{tabular}{c c | c c | c c}
\textbf{Partial} & & \textbf{Partial} & & \textbf{Partial} & \\
\textbf{Derivative} & \textbf{Units} & \textbf{Derivative} & \textbf{Units} & \textbf{Derivative} & \textbf{Units} \\ [0.5ex]
\hline
$ \partial a/\partial x_j $ & mm\textsuperscript{-1} & $ \partial b/\partial x_j $ & -- & $ \partial c/\partial x_j $ & mm \\
$ \partial a/\partial E_k $ & keV\textsuperscript{-1} & $ \partial b/\partial E_k $ & mm keV\textsuperscript{-1} & $ \partial c/\partial E_k $ & mm\textsuperscript{2} keV\textsuperscript{-1} \\
$ \partial z/\partial a $ & mm & $ \partial z/\partial x_j $ & -- & $ \partial \theta/\partial E_k $ & rad keV\textsuperscript{-1} \\
$ \partial z/\partial b $ & -- & $ \partial z/\partial E_k $ & mm keV\textsuperscript{-1} & -- & -- \\
$ \partial z/\partial c $ & mm\textsuperscript{-1} & -- & -- & -- & -- \\
\hline
\end{tabular}
\caption{Units for partial derivatives}
\end{table}

\par
The differences between the analytical and numerical results are generally within 0.5\%. Using our validated formulae, it's clear that the position resolution error component $ \sigma_{img,pos} $ dominates the total imaging error. Energy resolution only affects the Compton cone angle, and for this example we see $ \sigma_\theta = $ 0.00292 radians (0.168\textdegree{}). A final note -- with (3.000 mm, 2.000 keV) detector resolution, both the analytic and numeric estimated errors for this sample track are 29.253 mm. This relatively small imaging error can be attributed to the track's relatively long Compton lever arm of 34.387 mm. Analytic errors in the hundreds (or even thousands) of mm have been observed for other tracks where $ \bm{X_1} $ and $ \bm{X_2} $ are closer together.




\end{document}